\documentclass[twocolumn]{aastex631}
\usepackage{amsmath}

\shorttitle{AASTeX v6.3.1 Sample article}
\shortauthors{Liu et al.}
\graphicspath{{./}{figures/}}

\begin{document}
\title{VLBA reveals the absence of a compact radio core in the radio intermediate quasar J2242+0334 at $z=$5.9}

\correspondingauthor{Ran Wang}
\email{rwangkiaa@pku.edu.cn}

\author[0000-0001-9321-6000]{Yuanqi Liu}
\affiliation{Shanghai Astronomical Observatory, Key Laboratory of Radio Astronomy, CAS, 80 Nandan Road, Shanghai 200030, China}
\affiliation{Kavli Institute for Astronomy and Astrophysics, Peking University, Beijing 100871, P. R. China}

\author[0000-0003-4956-5742]{Ran Wang*}
\affiliation{Kavli Institute for Astronomy and Astrophysics, Peking University, Beijing 100871, P. R. China}

\author[0000-0003-3168-5922]{Emmanuel Momjian} 
\affiliation{National Radio Astronomy Observatory, P.O. Box O, Socorro, NM 87801, USA}

\author[0000-0001-8256-8887]{Yingkang Zhang}
\affiliation{Shanghai Astronomical Observatory, Key Laboratory of Radio Astronomy, CAS, 80 Nandan Road, Shanghai 200030, China}

\author[0000-0003-4341-0029]{Tao An}
\affiliation{Shanghai Astronomical Observatory, Key Laboratory of Radio Astronomy, CAS, 80 Nandan Road, Shanghai 200030, China}

\author[0000-0002-4439-5580]{Xiaolong Yang}
\affiliation{Shanghai Astronomical Observatory, Key Laboratory of Radio Astronomy, CAS, 80 Nandan Road, Shanghai 200030, China}

\author{Jeff Wagg}
\affiliation{SKA Observatory, Lower Withington Macclesfield, Cheshire SK11 9FT, UK}

\author[0000-0002-2931-7824]{Eduardo Ba\~nados}
\affiliation{ Max-Planck-Institut für Astronomie, Königstuhl 17, D-69117, Heidelberg, Germany}

\author[0000-0002-4721-3922]{Alain Omont}
\affiliation{Institut d’Astrophysique de Paris, Sorbonne Université, CNRS, UMR 7095, 98 bis bd Arago, 75014 Paris, France}

\begin{abstract}
High-resolution imaging is crucial for exploring the origin and mechanism of radio emission in quasars, especially at high redshifts. We present 1.5 GHz Very Long Baseline Array (VLBA) images of the radio continuum emission from the radio-intermediate quasar (RIQ) J2242+0334 at $z = 5.9$. 
This object was previously detected at both 1.5 GHz and 3 GHz with the Karl G. Jansky Very Large Array (VLA) as a point source. However, there is no clear detection in the VLBA images at both the full resolution of 10.7 milliarcsecond (mas) $\times$ 4.5 mas (61.7 pc $\times$ 26.0 pc) and a tapered resolution of 26 mas $\times$ 21 mas (150 pc $\times$ 121 pc). This suggests that the radio emission from the quasar is diffuse on mas scales with surface brightness fainter than the $3\sigma$ detection limit of 40.5 $\mu \rm Jy \ beam^{-1}$ in the full resolution image. The radio emission in the RIQ J2242+0334 is likely to be wind-like (i.e., diffuse) rather than in the form of collimated jets. This is different from the previous radio detections of the most luminous quasars at $z \sim$6 which are usually dominated by compact, high brightness temperature radio sources. Meanwhile, compared with RIQs at low redshifts, the case of J2242+0334 suggests that not all RIQs are beamed radio-quiet quasars. This optically faint RIQ provides an important and unique example to investigate the radio activity in the less powerful active galactic nuclei at the earliest cosmic epoch.

\end{abstract}


\section{Introduction} \label{sec:intro}
High-redshift quasars are ideal targets to explore the rapid accretion of supermassive black holes (SMBHs) and the nuclear activity at an early evolutionary stage \citep{Volonteri2012,Shen2019}. Currently, more than 300 quasars have been discovered at redshift $z\ge 5.7$, within the first 1 Gyr after the Big Bang (e.g., \citealt{Fan2000,Fan2001,Willott2010,Willott2010b, Jiang2016,Banados2016,Matsuoka2016,Matsuoka2018b,Wang2019}). 
The radio-loud fraction is constrained to be $10\%$ among these quasars \citep{Banados2018,Liu2021}. Previously, eight radio sources at $z\sim 6$ were imaged using Very Long Baseline Interferometry (VLBI) on scales of a few tens of pc to kpc (e.g., \citealt{Frey2011, Cao2014, Momjian2018, Momjian2021,Zhang2021}). 
The VLBI images of these high$-z$ quasars reveal compact radio cores and/or $\sim$ 1kpc scale radio jets, which are similar to the Compact/Medium-size Symmetric Objects (CSOs/MSOs) found at lower redshift \citep{Conway2002,An&Baan2012,Taylor2000}, arguing for a young radio-loud AGN at its early evolutionary stage.

These existing VLBI observations mainly focus on the most radio powerful objects at $z\sim 6$ with radio loudness $R>100$ ($R=f_{\rm radio}/f_{\rm optical}$), e.g., P352-15 with $R>1000$ \citep{Banados2018,Momjian2018}, and blazar PSO J0309+27 with $R=2500\pm500$ at $z\ge 6$ \citep{Belladitta2020}. Although it is limited by the sensitivity of the telescopes, studying much weaker sources is crucial to understanding the radio properties of the whole quasar population in the distant Universe.

In this paper, we present Very Long Baseline Array (VLBA) observations of a less luminous radio quasar CFHQS J224237+033421 (hereafter J2242+0334) at redshift 5.9, which is discovered in the Canada-France High$-z$ Quasar Survey (CFHQS, \citealt{Willott2010}). This quasar has a bolometric luminosity of $L_{\rm bol} = 4.2\times10^{12}L_{\odot}$, converted from the absolute magnitude of $-24.17$ at rest-frame 1450$\rm \AA$ \citep{Willott2010,Liu2021}. It is at the faint end of the quasar luminosity function which span a range from $10^{11}$ to $10^{14}L_{\odot}$ \citep{Wu2015,Matsuoka2018,Shen2019}.
As indicated by the quasar luminosity function, the relatively optically-faint quasars at $z\sim 6$ discovered from the deep optical and near-infrared surveys represent the more common quasar population in the early Universe \citep{Matsuoka2018}. 

Subsequent radio observation of J2242+0334 using the Karl G. Jansky Very Large Array (VLA) measures the radio loudness of $R = f_{\rm 5 GHz}/f_{\rm 4400\AA}= 54.9\pm 4.7$ \citep{Kellermann1989,Liu2021}. This value is above the boundary of $R$=10 for radio-loud quasars \citep{Kellermann1989}. In contrast to the bright radio sources with $R>$100, such objects with a less powerful radio source are usually called radio-intermediate quasars (RIQ,i.e., $3<R<100$) in the literature \citep{Miller1993,Diamond2009,Goyal2010}. Thus, J2242+0334 provides us a unique opportunity to investigate the early radio activity of radio-intermediate quasars.



J2242+0334 have been observed by VLA at both 3 GHz and 1.4 GHz \citep{Liu2021}. The VLA observation of J2242+0334 at 3 GHz and spatial resolution of $0.8^{\prime \prime}\times 0.7^{\prime\prime}$ (4.6 kpc $\times$ 4.0 kpc) revealed an unresolved radio source with a flux density $S_{\rm VLA-3GHz} = 87.0\pm 6.3 \rm \ \mu Jy$ at the optical quasar position. Observations at 1.4 GHz also reported an unresolved source with a flux density $S_{\rm VLA-1.4GHz} = 195.9\pm 24.7 \rm \ \mu Jy$ and a restoring beam size of $1.4^{\prime\prime} \times 1.2^{\prime\prime}$ (8.0 kpc $\times$ 6.9 kpc). Here, we present the VLBA observation of this object at 1.5 GHz with milliarcsecond (mas) resolution to further investigate the nature of the radio emission. The observations and data reduction are described in Section 2. We present the results in section 3, and the discussion in Section 4. Section 5 summarizes the conclusions.

Throughout this work, we adopt a $\rm \Lambda CDM$ cosmology with $\rm H_0 = 70 \ km \ s^{-1} Mpc^{-1}$, $\Omega_{\rm m} = 0.3$ and $\Omega_{\Lambda} = 0.7$. With this cosmological model, 1$\arcsec$ at the redshift of J2242+0334 corresponds to 5.77 kpc. 

\section{Observation and data reduction} \label{sec:data}
The VLBA observations (project code: BL280) of the quasar J2242+0334 were conducted at 1.5 GHz (L-band, 20 cm) during October $16^{th}$-$18^{th}$ in 2020. The observations were divided into three separate observing sessions. 
The optical coordinates (J2000) of the quasar are RA $=22^h42^m37.533^s$, dec$=+03\arcdeg34\arcmin22.03\arcsec$. The radio center coordinates in VLA 3 GHz images are RA$=22^h42^m37.5306^s$, dec$=+03\arcdeg34\arcmin22.0445\arcsec$, which are used as the target phase center in VLBA observations.
All ten VLBA stations were used. The total bandwidth, from 1.35 to 1.75 GHz, was composed of eight discrete 32 MHz intermediate frequency channels (IFs) in both right- and left-hand circular polarizations. Each channel was further split into 256 spectral points. The spectral points affected by Radio Frequency Interference (RFI) were flagged before calibration.
The data were recorded at each station with the ROACH Digital Backend and the polyphase filterbank (PFB) digital signal-processing algorithm with a data rate of 2048 Mbits/sec. Then, the data were correlated with 2s correlator integration time, using the VLBA DiFX correlator in Socorro, NM.

For the target, we employed a nodding-style phase referencing observing strategy with a series of 4-minute cycles, comprising of 3 minutes on the target and 1 minute on the phase calibrator. The total observing time was 12 hours, with 8.4 hours on target. We chose a bright and nearby phase calibrator J2245+0500 ($S_{\rm 1.5 GHz} = 153.3 \pm 1.2 \ \rm mJy$), which was 1.7 degrees away from the target. The source 3C454.3 was used as the fringe finder and bandpass calibrator.

The data were edited, and calibrated using the US National Radio Astronomy Observatory (NRAO) Astronomical Image Processing System (AIPS; \citealt{Greisen2003}) in a standard way. Imaging and the \texttt{CLEAN} task of the target and calibrators were applied with the Caltech Difmap (\citealt{Shepherd1997}). Then we performed self-calibration for both phase and amplitude on the phase calibrator J2245+0500 based on its CLEAN model. The self-calibration resolutions were also applied to the target visibility data. The main parameters of the VLBA observations are summarized in Table \ref{table1}.

\begin{table}[pt]
\tabletypesize{\scriptsize}
\caption{observation parameters}
\begin{tabular}{llll}
\hline
\hline
Observing central frequency (GHz)  & 1.5 \\ 
Recorded bandwidth (MHz) & 256 \\
Date of observations (in 2020) &$16^{th}-18^{th}$ Oct \\
Phase calibrator   &J2245+0500    \\  
Flux/Bandpass calibrator   &  3C454.3    \\ 
\hline 
Full resolution image: & \\
FWHM beam size* (mas) & 10.7$\times$4.5 \\ 
Beam P.A. ($^{\circ}$) &-2.8 \\
Image $1\sigma$ rms sensitivity ($\mu \rm Jy \  beam^{-1}$) &13.5 \\
Brightness temperature (K) & $\leq 3\times 10^6$ \\

\hline 
Tapered image at 4M$\lambda$: & \\
FWHM beam size* (mas) & 26$\times$21 \\ 
Beam P.A. ($^{\circ}$) &22.0 \\
Image $1\sigma$ rms sensitivity ($\mu \rm Jy \  beam^{-1}$) & 32.6 \\
\hline  
\end{tabular}
\renewcommand\arraystretch{1}
\begin{flushleft}
*note: Natural weighting.
\end{flushleft}
\label{table1}
\end{table}
\begin{figure*}[t]
\gridline{\fig{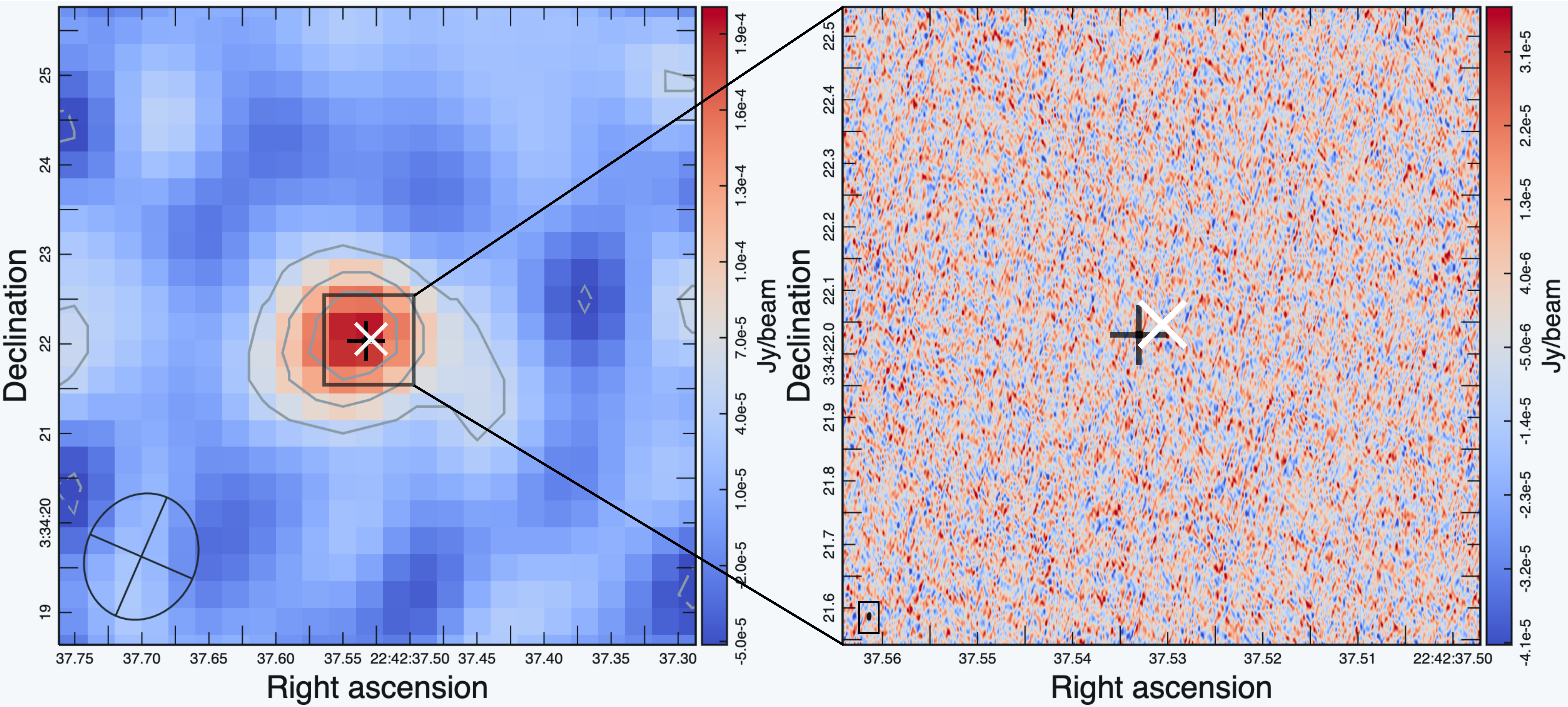}{1\textwidth}{}}
\caption{Left Panel: The intensity map of VLA observations at 1.4 GHz. The contour levels are at $[-$1, 2, 4, 6$]$ times the corresponding $1\sigma$ RMS noise level of $24.7 \ \rm \mu Jy/beam$. The synthesized beam with FWHM of $1.4\arcsec \times 1.2\arcsec$ is shown as an ellipse in the bottom left corner. 
Right Panel: Full resolution VLBA image in $1^{\prime \prime}\times 1^{\prime \prime}$ area, zooming into the central area in the VLA image. A synthesized beam with FWHM of 10.7 mas$\times$4.5 mas is shown as the ellipse in the bottom left black square box. 
For both panels, the black plus denotes the optical position and the white cross represents the peak position in VLA 3 GHz observations \citep{Liu2021}, which is also the phase center of the VLBA observations. 
\label{figure1}}
\end{figure*}


\section{Results and analysis} \label{sec:results}

We first image the VLBA data at its full resolution and natural weighting. The full width at half maximum (FWHM) beam size is 10.7 $\times$4.5 mas (61.7 pc $\times$ 26.0 pc at $z=5.9$) and the $1\sigma$ rms is 13.5 $\mu \rm Jy \ beam^{-1}$ (Figure \ref{figure1}).  No radio source is detected at the optical quasar or VLA radio source position, which are $0.4\arcsec$ apart. This results in a $3\sigma$ upper limit of 40.5 $\mu \rm Jy \ beam^{-1}$ for the surface brightness of the radio source. 
This estimates the intrinsic brightness temperature to be $T_b \leq 3 \times 10^6 \ \rm K$ for any compact radio source at the quasar position.
Here we estimate the $3\sigma$ upper limit of the intrinsic brightness temperature adopting the equation in \citet{Synthesis1999}:
\begin{equation} \label{eq1}
T_b = 1.22\times 10^{12}\left(1+z\right) \frac{S_{\nu}}{\nu^2 \theta_{\rm max}\theta_{\rm min}},
\end{equation}
where $\nu$ is the observing frequency in GHz, $S_\nu$ is the surface brightness detection limit in $\rm Jy \ beam^{-1}$, and the resolution $\theta_{\rm max} = 10.7$ mas and $\theta_{\rm min}=4.5$ mas, which is converted from the Rayleigh-Jeans law ($T_b = \frac{\lambda^2}{2k\Omega}S$, $k$ is the Boltzmann constant, $\Omega$ is the beam solid angle).


To reduce the resolution and recover more extended emission, we apply a two-dimensional Gaussian taper to the UV visibililty data, which reduces the weight to $50\%$ at UV distance $4M\lambda$ (in unit of wavelength). The resolution of the tapered image is 26 mas $\times$ 21 mas (150 pc × 121 pc). 
No $>5\sigma$ signal is detected in the source radio position and surrounding 5 kpc area.  

\section{Discussions} \label{sec:dis}
The non-detection of J2242+0334 in the VLBA image, together with the previous VLA detection, constrains the spatial extent of the radio emission from this object. The source is unresolved with the VLA in B-array configuration \citep{Liu2021} with a minimum resolvable size to be $\theta = b\sqrt{\rm{\frac{4ln2}{\pi}ln(\frac{SNR}{SNR-1})}} =0.4^{\prime\prime} $, where $b=\sqrt{b_{max}\times b_{min}} =1.3^{\prime\prime}$ is the beam size \citep{Kovalev2005}. This minimum resolvable size corresponds to 2.3 kpc at the redshift of J2242+0334. 
The non-detection in both the full-resolution and the tapered VLBA images suggest that the emission detected by the VLA is not from a compact source at the mas-scale resolution. The radio emission is diffuse on tens of mas to sub-arcsec (tens of pc to kpc) scales. 

The 1.4 GHz and 3 GHz data presented in \citet{Liu2021} show a steep power-law spectrum with a spectral index $\alpha_R = -1.07^{+0.27}_{-0.25}$ (adopting the model $S_{\nu} \propto \nu^{\alpha}$).
This steep spectrum indicates that the diffuse radio emission detected in J2242+0334 is likely to be dominated by optically-thin synchrotron emission \citep{Silpa2020,Odea2021,Patil2022}. Such optically-thin synchrotron emission with resolved/extended morphology and steep spectra was typical for radio-quiet quasars observed at low redshifts (e.g., \citealt{Pei2020}).

The radio emission, although diffuse, is unlikely to be powered primarily by nuclear star formation.
If the VLA detected radio emission $S_{\rm VLA-1.4GHz} = 195.9\pm 24.7 \rm \ \mu Jy$ is all from star formation, the star formation rate (SFR) would be $\sim 55000 \ \rm M_{\odot}yr^{-1}$, according to the relation in \citet{Murphy2011}:\begin{equation} \label{eq1} \left(\frac{\rm{SFR_{1.4 \ GHz}}}{ \rm M_{\odot} \ yr^{-1}}\right) = 6.35 \times 10^{-29} \left(\frac{L_{\rm 1.4 \ GHz}}{\rm erg \ s^{-1} Hz^{-1}}\right) , \end{equation} where the rest-frame radio luminosity at 1.4 GHz is extrapolation with the spectral index measured by VLA.
However, J2242+0334 was not detected by $Herschel$ in bright dust continuum in submillimeter Bands. \citet{Lyu2016} constrained the SFR upper limit of J2242+0334 to be $2561 \ \rm M_{\odot}yr^{-1}$ from the integrated infrared (IR) luminosity upper limit. The difference between SFRs converted from radio and the IR detection limit suggests that star formation cannot be the dominant power source of the detected radio emission with the VLA. Thus, most of the radio emission from J2242+0334 is likely to be powered by the AGN.  One possibility is that the VLA-detected radio flux density consists of winds or several knots with surface brightness below the detection limit. Furthermore, a less powerful jet may interact with the surrounding medium, which could also result in extended radio emission. Such systems were identified in some local quasars with radio and X-ray observations \citep{Wills1999, Kunert2010, Silpa2021}.  


\subsection{Comparison with quasars at $z\sim 6$}



Currently, among more than 300 quasars found at $z>5.7$, only eight (7 radio-loud and 1 radio-quiet) quasars have been observed with the VLBI technique before this work. Among these, J2242+0334 is the first source that no compact core is detected at mas-scale resolution, despite being comparable or even deeper VLBI observations.  

All the previous seven radio-loud quasars are observed in the form of compact structures at the mas scale, showing as a radio core (e.g., P172+18, \citealt{Momjian2021}; J2318-3113, \citealt{Zhang2021}) or multiple components (one-side jet or CSOs, e.g., J1427+3312, \citealt{Frey2008}; PSO J352.4034-15.3373, \citealt{Momjian2018}). They are reported with high intrinsic brightness temperatures of $10^6-10^9 \rm K$ and relatively steep spectra ($\alpha = -1.1 \sim -0.8$). These radio-loud sources appear to be the early stage of powerful radio galaxies.

Comparing the VLA and VLBI observations at the same frequency, at least 2 quasars (SDSS J2228+0110, \citealt{Cao2014}; PSO J352.4034-15.3373, \citealt{Momjian2018}) report significant flux density differences ($>50\%$).
The differences between VLA and VLBA for these two objects indicate the radio emission from both compact core and extended component (likely to be winds) on kpc scale. Compared to these radio-loud quasars, J2242+0334 seems to be dominated by diffuse, large scale emission along with likely small-scale emission too faint to be detected in our observations.

The only radio-quiet ($R=0.6$) quasar at $z > 6$ with VLBI observations so far, J0100+2802 shows an unresolved radio core ($88\pm 19 \ \rm \mu Jy$) in the VLBA image at $\sim 10$ mas resolution \citep{Wang2017}. The compact core accounts for about $65\%$ of the flux density detected by the VLA at the same frequency ($136.2 \pm 10.4 \ \rm \mu Jy$) observed in the same year \citep{Liu2022}. J2242+0334 and J0100+2802 have a comparable radio flux density in the VLA observations \citep{Wang2016}. However, they are quite different on mas scales as J2242+0334 shows no compact radio source, while J0100+2802 shows a compact source with high brightness temperature $T_b=(1.6 \pm 1.2)\times 10^7$K and relatively flat spectrum with the power-law index of $\alpha = -0.52 \pm 0.18$ fitted from VLA observations at 1.5 GHz, 6 GHz, and 10 GHz. 

Compared to J0100+2802 with a bolometric luminosity of $L_{\rm bol} = 4.29 \times 10^{14} L_{\odot}$ \citep{Wu2015}, which hosts the most massive and luminous AGN, the nucleus of J2242+0334 is much less powerful with a bolometric luminosity about 2 orders of magnitude smaller. Although there is radio activity from both nuclei with comparable radio flux density, the jet from J2242+0334 may be less energetic and/or uncollimated. Such an uncollimated jet or wind-like structure may interact with the surrounding medium with a large impact factor, resulting in diffuse radio emission.

\subsection{Comparison with low-redshift RIQs}

The RIQ population at low redshifts were observed and studied at both VLA and VLBI scale, to explore the nature of their radio activity and to understand how they compare to the radio-quiet and radio-loud objects. A substantial fraction of RIQs is suggested to be relativistically beamed counterparts of weak jets in radio-quiet quasars, based on their observational features of core-dominated morphology (a few have extended emissions) with flat radio spectra and optical variability \citep{Falcke1996, Wang2006, Goyal2010}. However, the beaming scenario is not conclusive from the study of 11 RIQs at redshift 2-4 with the European Very-Long-Baseline-Interferometry Network (EVN) observations \citep{klockner2009}.
Among the sample of 11 RIQs, compact cores are detected in 7 sources while $\sim 40\%$ of the VLA measured flux densities are resolved out on scale larger than 150 pc. Therefore, not all of these quasars can be explained as beamed radio-quiet quasars. The fraction of the emission resolved out by EVN can be related to the size and orientation of the jet. In the case of J2242+0334, the radio emission is totally resolved out in the VLBA image. Therefore, it is unlikely to be a beamed source.


The best-studied RIQ, III Zw 2, is highly variable, containing complex components at kpc scales, including jets, lobes and uncollimated winds \citep{Brunthaler2000,Brunthaler2005}. 
The polarization study indicates that the weak jet in this RIQ is suppressed by the surrounding strong wind-like component. Radio variability and polarization structures also indicate that the jet activity is intermittent, which may be explained by changing spectral states of the accretion disk in an ongoing merger event \citep{Silpa2021}. 
Our VLBA observation of J2242+0334 reveals similar radio emission properties with III Zw 2, where the diffuse emission is the dominant component for the emission detected with the VLA. It is possible that J2242+0334 is the counterpart of the RIQ systems like III Zw 2 at the highest redshift. Further monitoring of the radio variability and accurate measurements of the BH mass and X-ray luminosity will help to explore the evolutionary properties of the central engine e.g., intermittent activity, changes of accretion disk states, and surrounding magnetic field influence.



\section{summary}
We carried out VLBA observations on the radio-intermediate quasar J2242+0334 at $z=5.9$. The source is undetected in both full resolution and tapered images. We here provide a point source $3\sigma$ detection limit of 40.5$\mu \rm Jy \ beam^{-1}$ at 10.7 mas $\times$ 4.5 mas resolution. The brightness temperature is constrained to be $T_b\leq 3 \times 10^6$K.
This is the first time we report diffuse radio emission from a RIQ at $z\sim6$ which is fully resolved out at the resolution of VLBA. Such extended radio emission with low surface brightness is likely dominated by weak (undetected) jet knots and/or originates from the interaction between the weak jet and the surrounding medium. It also indicates that not all RIQs are beamed radio-quiet quasars. Thus J2242+0334 makes a unique example for the complex radio activity in the RIQ in the early Universe. 


\begin{acknowledgments}
We acknowledge the supports from the National Science Foundation of China (NSFC) grants No.11721303, 11991052, 11373008, 11533001. The National Radio Astronomy Observatory (NRAO) is a facility of the National Science Foundation operated under cooperative agreement by Associated Universities, Inc. This paper makes use of the VLBA data from program BL280.
\end{acknowledgments}

\facilities{VLBA, AIPS, Difmap}



\bibliography{sample631.bib}{}
\bibliographystyle{aasjournal}

\end{document}